\documentclass[12pt]{article}
\usepackage[T1]{fontenc}
\usepackage[francais,english]{babel}
\usepackage[dvips]{graphicx}
\usepackage{amsfonts}
\usepackage{latexsym}
\usepackage{mathptmx}
\usepackage{amsmath}
\usepackage{lscape}
\usepackage{longtable}
\usepackage{lscape}
\usepackage{rotating}
\usepackage{longtable}
\usepackage{setspace}
\usepackage{anysize}

\marginsize{.9in}{.9in}{1.5in}{1.5in}

\def\picture #1 by #2 (#3){\dimen0 = \hsize
\advance\dimen0 by -#1 \divide\dimen0 by 2 \hskip
\dimen0 \vbox to #2{\hrule width #1 height 0pt depth 0pt
\vfill \special{picture #3}}}

\def \R{{\mathbb R}}

\font\ninerm=cmr9
\long\outer\def\abstract#1{\bigskip\vbox{\noindent\ninerm
\baselineskip=10pt#1}\nobreak\bigskip}
\catcode`\â= \active \def â{\^a}
\catcode`\ê= \active \def ê{\^e}
\catcode`\ô= \active \def ô{\^o}
\catcode`\û= \active \def û{\^u}
\catcode`\î= \active \def î{{\^\i}}
\catcode`\é= \active \def é{\'e}
\catcode`\è= \active \def è{\`e}
\catcode`\à= \active \def à{\`a}
\catcode`\ù= \active \def ù{\`u}
\catcode`\ç= \active \def ç{\c {c}}
\catcode`\ï= \active \def ï{{\"\i}}

\newcount\numero
\def\exo#1{\advance\numero by 1\bigskip
{\noindent\tenbf #1\the\numero. }}
\def\frac#1#2{{#1\over #2}}
\numberwithin{equation}{section}

\begin{document}

\title{Historical risk measures on stock market indices and energy markets}
\author{Wayne Tarrant\footnote{Department of Mathematics, Wingate University, Wingate, NC 28174, USA, e-mail: w.tarrant@wingate.edu, +1 (704) 233-8189}}

\maketitle

\vfil \eject

\begin{abstract}

\noindent \textit{Abstract}: In this paper we look at the efficacy of different risk measures on energy markets and across several different stock market indices. We use both the Value at Risk and the Tail Conditional Expectation on each of these data sets. We also consider several different durations and levels for historical risk measures. Through our results we make some recommendations for a robust risk management strategy that involves historical risk measures. \\

\noindent \textit{Keywords}: Historical risk measure - Value at Risk - Risk management - energy markets - stock market indices - Tail Conditional Expectation \\

\noindent \textit{JEL}: C16 - G18 - E52 \\

\end{abstract}

\vfil \eject

\section{Introduction}

\noindent Concepts of risk have always been important in financial markets. From Markowitz's use of variances and covariances to later uses of semivariance and beta through to the modern measures of Value at Risk and Tail Conditional Expectation, academics and practitioners have long been trying to put a number on the amount of risk that certain positions entail. The final two measures have found their way into much current regulation, so one would assume a great deal is known about each. The amount that is unknown is astonishing. \\

\noindent In this paper we will introduce the modern concepts of risk measures. We will give the reasoning behind certain properties of risk measures and give definitions of the two most widely used risk measures. We will then show errors that are encountered if we choose to use an individual risk measure on historical returns as a sole predictor, just as is required by the Basel Accords. We will demonstrate these difficulties by looking at data on stock market indices and energy markets. Finally, we will make a suggestion about how to use multiple risk measures in order to have a better view of the entire risk profile of a position over many durations. \\

\section{Risk Measures}

\noindent Some have objected to a risk measure being a single number, but there is some support for this idea. Investing is always a binary decision- either one invests or one chooses not to invest. Thus the argument is that, given a single number, one should have enough information to decide whether to invest or not. There have been some general agreements about the kinds of properties that such a risk measure ought to possess. These agreements must be acknowledged as being assumptions, but they have provided the definition that is now universal.\\

\noindent We recall the universally accepted definition of a risk measure (Artzner et al., 1997). Let $X$ be a random variable. Then $\rho$ is a risk measure if it satisfies the following properties:
\begin{itemize}
\item (Monotonicity) if $X \ge 0$, then $\rho(X) \le 0$
\item (Positive Homogeneity) $\rho(\alpha X) = \alpha \rho(X) \ \forall \ \alpha \ge 0$
\item (Translation Invariance) $\rho(X + a) = \rho(X) - a \ \forall \ a \in \R$.
\end{itemize}

\noindent The idea behind this definition is that a positive number implies that one is at risk for losing capital and should have that positive number of a cash balance on hand to offset this potential loss. A negative number would say that the company has enough capital to take on more risk or to return some of its cash to other operations or to its shareholders. The monotonicity property states that an investment that always has positive payoff gives the company the ability to take on more risk. Positive homogeneity implies that multiplying your investment by $k$ times gives you a risk of a loss that is $k$ times larger. Translation invariance implies that a company holding $a$ in cash lowers its measure of risk by $a$.\\

\subsection{Value at Risk}

\noindent The most well-known risk measure is the Value at Risk (VaR). In order to define VaR, we must first recall some basic concepts.
Let $X$ be a random variable and $\alpha \in \lbrack 0,1 \rbrack $.
\begin{itemize}
\item  $q$ is called an \bf{$\alpha$-quantile} \rm if $Pr \lbrack X < q \rbrack \le 1 - \alpha \le Pr \lbrack X \le q \rbrack$,
\item the largest \bf{$\alpha$-quantile} \rm is $q_{\alpha}(X) = inf \lbrace x | Pr \lbrack X \le x \rbrack > 1 - \alpha \rbrace$, and
\item the smallest \bf{$\alpha$-quantile} \rm is $q_{\alpha}^- = inf \lbrace x | Pr \lbrack X \le x \rbrack \ge 1 - \alpha \rbrace$.
\end{itemize}

\noindent It is easy to show that $q_{\alpha} \ge q_{\alpha}^-$ and that $q$ is an $\alpha$-quantile if and only if $q_{\alpha}^- \le q \le q_{\alpha}$.\\

\noindent Given a position $X$ and a number $\alpha \in \lbrack 0,1 \rbrack$, we define the $\alpha$-Value at Risk,  $VaR_{\alpha}(X)$, by $VaR_{\alpha}(X) = -q_{\alpha}(X)$. We call $X$ $\alpha$-VaR acceptable  if $VaR_{\alpha}(X) \le 0$ or, equivalently, if $q_{\alpha}(X) \ge 0$. The $\alpha$-VaR can be seen as the amount of cash that a firm needs in order to make the probability of that firm going bankrupt to be equal to $\alpha$. This leads to the following property of VaR: $VaR_{\alpha}(X + VaR_{\alpha}(X)) = 0$. This states that one may offset the risk of an investment by having an amount of cash on hand equal to the Value at Risk inherent in holding the asset.\\

\noindent The Value at Risk has historically been the most widely used of the risk measures. For instance, it is the measure used by the Basel Committee on Banking Supervision in the Basel I and Basel II accords (BCBS 1988, BCBS 1996) in order to determine capital requirements for banks. However, there are two major problems cited with Value at Risk as a risk measure.\\

\noindent  The issue that is easier to describe is the fact that Value at Risk does not address how large of a bankruptcy a firm will experience. This risk measure only tells that a business will be bankrupt with a 5\% chance if it doesn't have on account a sum of money equal to the 95\%-VaR.
As an example, say that a firm holds an asset that initially had no cost. That asset has a 95\% chance of paying the company US\$2 million and a 5\% chance of giving the company a loss of US\$1 million. By the above definition, we find that the 95\%-VaR is US\$1 million.
Now assume that the firm holds another asset, again without initial cost. This asset again has a 95\% chance of giving the company a profit of US\$2 million, with a 1\% chance of a loss of US\$1million and a 4\% chance of a loss of US\$1 billion. This asset also has a 95\%-VaR of US\$1 million, but it is very clear that the second asset is much more risky for the firm to hold.\\

\noindent The second problem critics cite is that VaR does not account for diversification effects. It is widely held that owning different types of investment vehicles will help to alleviate some of the risk that holding only one investment would produce. (Markowitz 1952) discusses diversifying away specific risk, that risk that is unique to a given holding. He acknowledges that one cannot alleviate the market risk that is common to all holdings. Markowitz's calculations suggest that holding a sufficiently large number of assets, chosen with appropriate covariance, diversifies away specific risk.\\

\noindent As an example of this problem with VaR, consider the case where a bank has made two \$1 million loans and one \$2 million loan, each with a 0.04 probability of default and all pairwise independent. Then the 95\%-VaR for each loan is \$0. Thus, if we construct a portfolio consisting solely of the \$2 million loan, we must have a 95\%-VaR of \$0. If we instead choose diversification and make our portfolio out of the two independent \$1 million loans, something paradoxical happens. The probability of both loans defaulting is 0.0016, but the probability of exactly one loan defaulting is 0.0768. This implies that the 95\%-VaR of our diversified portfolio is \$1 million. Thus, VaR does not favor diversification in some instances.\\

\noindent  So, is diversification always the best policy? If each of our assets has some probability of default, is it better to spread the risk of loss over several loans, choosing a larger probability of taking a small loss? Or does it make more sense to hold only one asset, taking a small probability of losing everything? These are questions of risk aversion that each bank and each individual investor will need to answer depending on his or her or its own risk preferences. Investment professionals have decided that the former is preferable to the potential of losing their entire investment. And some regulating agencies have agreed with this judgment, though the Basel Committee has implicitly disagreed, whether intentionally or not. Thus, we will pursue ways in which we might reconcile these differences and determine if there is some risk measure that might respect the desirability of diversification.

\subsection{Coherent Risk Measures}

In 1997 Artzner, Delbaen, Eber, and Heath (1997) (henceforth ADEH) proposed four desirable properties that measures of risk should have. In addition to the properties of translation invariance, positive homogeneity, and monotonicity, these authors added the property of subadditivity, i.e., using the notation from before,
$$
{\rm if} \; \; X_1, X_2\;  {\rm are}\;  {\rm random}\; {\rm variables,}\; {\rm then} \; \;  \rho(X_1 + X_2) \le \rho(X_1) + \rho(X_2).
$$
\noindent This is an attempt to respect the perceived value of portfolio diversification. They chose to call any measure that satisfied all four of the properties by the name ``coherent risk measure''. Of course this was an intelligent choice of names for their class of measures, as nobody would want to be accused of using a measure that is incoherent.\\

\noindent ADEH (1997) start by defining a set of acceptable positions. These are positions that some sort of supervisor, such as a regulator or an exchange's clearing firm, decides have an appropriate level of risk for the firm. They then propose that a firm must make choices when faced with an unacceptable position. Either the firm may alter the position or decide to offset the unacceptable position by having cash on hand to account for the possibility of a loss. Should the firm decide to hold the unacceptable position, a risk measure will be the ``distance'' away from acceptable that their position measures.\\

\noindent ADEH (1997) then go about showing links between acceptance sets and measures of risk. Their acceptance sets are those for which the risk measure $\rho$ takes a negative value on the set. Thus, acceptance sets are those for which no additional capital is needed to offset the risk that is being taken by holding the given position.\\

\noindent They next demonstrate that VaR is incoherent and also that VaR fails to recognize when there is a concentration of risks. This is just as we have shown above, although their examples are slightly more complicated and less transparent. The main conclusions of these authors  to reject Value at Risk as a good risk measure  lies on the following two reasons:
\begin{itemize}
\item  Value at Risk does not behave nicely with respect to addition of risks, even independent ones, creating severe aggregation problems.
\item the use of Value at Risk does not encourage and, indeed, sometimes prohibits diversification, because Value at Risk does not take into account the economic consequences of the events the probabilities of which it controls.
\end{itemize}

\noindent The authors (ADEH, 1997) then give the above definition of a coherent risk measure. They prove that any coherent risk measure arises as the supremum of the expected negative of final net worth for some collection of ``generalized scenarios'' or probability measures on states of the world. They point out that ``Casualty actuaries have been working long at computing pure premium for policies with deductible, using the conditional average of claim size, given that the claim exceeds the deductible, (Hogg and Klugman, 1984). In the same manner, reinsurance treaties
have involved the conditional distribution of a claim for a policy (or of the total
claim for a portfolio of policies), given that it is above the ceding insurer's retention
level. In order to tackle the question of \it{how} \rm bad is bad, which is not addressed by
the Value at Risk measurement, some actuaries (Bassi, Embrechts and Kafetzaki, 1996) have first identified
the deductible (or retention level) with the quantile used in the field of financial
risk measurement.'' \\

\noindent We now introduce the notion of tail conditional expectation.  Given a base probability
measure P on a probability space $\Omega$, and a level $\alpha$, the tail conditional expectation is the measure of risk defined by
\begin{equation} \label{tce}
TCE_{\alpha}(X) = -E_P [X | X \le - VaR_{\alpha}(X)]
\end{equation}

\noindent This concept is variously called the expected shortfall, expected tail loss, conditional value at risk, mean excess loss or mean shortfall by several authors in the literature (Rockafellar  and Uryasev 2002).
(ADEH) are able to show that coherent risk measures in general (and so their tail conditional expectation in particular) correspond to convex closed sets which include the positive orthant and are in fact homogeneous cones. They give several classifications, showing that coherent risk measures are in correspondence with the measures that arise from looking at the ``generalized'' scenarios that banks may face in order to construct and maintain acceptable positions. In mathematical terms, these are measures on probability spaces. \\

\noindent Finally, we can remark that for each risk $X$ one has the equality
$$VaR_{\alpha}(X) = inf \lbrace \rho(X) | \rho \; \; {\rm coherent}, \; \; \rho \ge VaR_{\alpha}(X) \rbrace.$$
Thus, knowing that
 more restrictive measures are available to them, the question is why regulators would use the Value at Risk, noting that no known organized exchanges use VaR as the basis of risk measurement for margin requirements. ADEH (1997) immediately answer their own question, with a quote from Stulz (1996), ``Regulators like Value at Risk because they can regulate it.'' \\

\section{Risk measures on energy markets and on stock indices}

\noindent For this study, we investigated the Brent, West Texas Intermediate (WTI), gasoline (GO), and heating oil (HO2) returns from 2 January 1995 to 11 March 2010. We also considered the French CAC-40 index from 9 July 1987 to 6 October 2009, the German DAX and the American Dow Jones 30 Industrials and the American S\&P 500 from 2 January 1980 to 6 October 2009. For each series of returns, we attempted to calculate the historical VaR and TCE at the 10 day- 90\% level, 20 day- 90\% level, 20 day- 95\% level, 50 day- 90\% level, 100 day- 90\% level, 100 day- 95\% level, 100 day - 99\% level, 250 day- 90\% level, 250 day- 95\% level, 250 day- 99\% level, 500 day- 90\% level, 500 day- 95\% level, and the 500 day- 99\% level. We were unable to calculate several of the TCEs because they did not exist at several durations and levels. Because many of the historical n-day time periods to be considered had no violations of the n-day historical VaR at some level $\alpha$, the n-day, $\alpha$ TCE did not exist, because division by zero would be required to compute that n-day, $\alpha$ TCE. The rates of nonexistence of the n day- $\alpha$ TCE for each of our assets are demonstrated in the next table. \\

\begin{table}

\centering

\begin{tabular}{|c|c|c|c|c|c|c|c|c|}
  \hline
  asset & Brent & CAC40 & DAX & Dow & GO & HO2 & S\&P & WTI \\
  10,90\% & .301 & .264 & .269 & .277 & .282 & .273 & .425 & .287 \\
  20,90\% & .072 & .055 & .050 & .060 & .062 & .058 & .061 & .072 \\
  20,95\% & .320 & .345 & .340 & .343 & .348 & .323 & .336 & .326 \\
  50,90\% & .006 & .020 & .010 & .008 & .005 & 0 & .007 & .002 \\
  100,90\% & 0 & 0 & 0 & 0 & 0 & 0 & 0 & 0 \\
  100,95\% & .054 & .025 & .017 & .022 & .014 & .004 & .023 & .001 \\
  100,99\% & .289 & .379 & .433 & .342 & .356 & .355 & .342 & .299 \\
  250,90\% & 0 & 0 & 0 & 0 & 0 & 0 & 0 & 0 \\
  250,95\% & 0 & 0 & 0 & 0 & 0 & 0 & 0 & 0 \\
  250,99\% & .128 & .148 & .151 & .126 & .197 &.181 & .093 & .191 \\
  500,90\% & 0 & 0 & 0 & 0 & 0 & 0 & 0 & 0 \\
  500,95\% & 0 & 0 & 0 & 0 & 0 & 0 & 0 & 0 \\
  500,99\% & 0 & .113 & .083 & .023 & .022 & .048 & .041 & 0 \\
  \hline
\end{tabular}

\caption{rates of TCE nonexistence}

\end{table}

\noindent From the table it is apparent that the TCE does not exist for smaller time periods. It also fails to exist if we want the level to be high. In fact the TCE only exists for the 100 day- 90\% level, 250 day- 90\% level, 250 day- 95\% level, 500 day- 90\% level, and 500 day- 95\% level. So it is quite clear that the TCE will not work for smaller timeframes or for high levels of certainty. \\

\noindent Since we don't have a TCE for the shorter timeframes, we can look to the VaR for these instances. When used as a predictor, the $\alpha$-VaR is a prediction that the returns will be below that $\alpha$-VaR exactly $1 - \alpha$ of the time. That is to say that we should see returns below the 95\% VaR approximately 5\% of the time. So, we took a look at the VaR on each of the markets over several different timeframes and at several different levels.  As we can see from Table 2, the VaR tends to have smaller errors for smaller timeframes, while the errors increase when the timeframe increases. We can also see increasing error rates as the level increases.\\

\begin{table}

\centering

\begin{tabular}{|c|c|c|c|c|c|c|c|c|}
  \hline
  asset & Brent & CAC40 & DAX & Dow & GO & HO2 & S\&P & WTI \\
  10,90\% & -.07692 & -.01607 & -.04412 & -.06233 & -.05897 & -.04103 & -.03979 & -.06923 \\
  20,90\% & -.07712 & -.0161 & -.04418 & -.06242 & -.05913 & -.03856 & -.04117 & -.06941 \\
  20,95\% & -.01538 & -.01071 &.032086 & -.03979 & -.06154 & -.02564 & .015915 & -.02051 \\
  50,90\% & -.00259 & .044964 & .00672 & -.0067 & -.03886 & -.02073 & .009383 & -.04663 \\
  100,90\% & .002625 & -.00181 & .028417 & .022942 & -..01575 & 0 & .009447 & .026247 \\
  100,95\% & .041885 & .141818 & .043243 & -.01348 & .031414 & .04712 & .018868 & .078534 \\
  100,99\% & .157895 & .036364 & .040541 & .054054 & 0 & .078947 & .121622 & .131579 \\
  250,90\% & .060109 & .018657 & .042818 & .004132 & .046448 & .051913 & .015152 & .04918 \\
  250,95\% & .120219 & .130597 & .140884 & .110193 & ..153005 & .10929 & .006116 & .071038 \\
  250,99\% & .189189 & -.01852 & -.08333 & -.05479 & -.13514 & -.08108 & .041096 & 189189 \\
  500,90\% & .064516 & .084149 & .071531 & .021398 & .067449 & .082111 & .028531 & .055718 \\
  500,95\% & .105263 & .017578 & .082857 & .088571 & .146199 & .076023 & .111429 & .070175 \\
  500,90\% & .147059 & .392157 & .3 & .414286 & .176471 & .147059 & .457143 & .352941 \\
  \hline
\end{tabular}

\caption{VaR prediction errors}

\end{table}

\noindent Note that in this chart a negative error actually means that the VaR was a conservative predictor of returns, meaning that the returns were smaller than the VaR less than the predicted number of times. A positive error here means that the VaR was not conservative enough. It implies that we saw returns that were smaller than the prediction more than the predicted number of times. Of course this is very dangerous, as an error of 10\% for the 90\% VaR means that the losses on that market were below the 90\% VaR 11\% of the time and not the predicted 10\%. This will certainly affect capitalization levels and could play a part in a number of bankruptcies if companies hold reserves right at the VaR's suggestion, as is often the case.\\

\noindent Regressing the VaR error rate over the duration and level, we find that the multiple Rs for all eight of the markets we have considered have a mean of 0.731. The multiple Rs range from 0.599 for the CAC40 up to 0.894 for WTI. P-values showed much more significance on time than on level. Taken as a group, the market indices have a multiple R of 0.572, with the P-value for time at 0.0001 and for level at 0.025. The oil markets have a multiple R of 0.729 with P-values of 0.000003 and 0.0007 for time and level, respectively. Statistically, shorter time durations yield much better predictions for the VaR. Further, the 90\% VaR is a much better predictor than the 95\% VaR or the 99\% VaR.\\

\noindent For the longer durations, for which we have just seen that the VaR is least useful, we must look at the TCE. As shown in Table 3, the actual returns on the TCE are very close to the returns that would be predicted by the historical TCE. In fact, where the TCE exists, the predictions are never off by more than 0.5\%. In the chart below, we note that a negative error means that the TCE prediction was not conservative enough. That is, the actual returns were lower than what was predicted.\\

\begin{table}

\centering

\begin{tabular}{|c|c|c|c|c|c|c|c|c|}
  \hline
  asset & Brent & CAC40 & DAX & Dow & GO & HO2 & S\&P & WTI \\
  100,90\% & -.002596 & -.001821 & -.001749 & -.001396 & -.00210 & -.002490 & -.001451 & -.002142 \\
  250,90\% & -.00167 & .000113 & -.00166 & -.00156 & -.00174 & -.0012 & -.0017 & -.00277 \\
  250,95\% & -.00302 & -.00295 & -.00253 & -.00227 & -.00218 & -.00214 & -.00256 & -.00456 \\
  500,90\% & -.001586 & -.002418 & -.002377 & -.002124 & -.001526 & -.001457 & -.00250 & -.002035 \\
  500,95\% & -.002489 & -.004112 & -.003462 & -.00355 & -.001915 & -.001491 & -.004008 & -.003529 \\
  \hline
\end{tabular}

\caption{TCE prediction errors}

\end{table}

\noindent A quick perusal of the chart shows that the TCE is most effective at the 250 day-90\% level. It is somewhat surprising that the 500-day level does not provide more accurate estimates in the energy markets. It is notable that only the 100 day-90\% level of VaR has smaller absolute error than the 250 day-90\% level of VaR. Of course this is reasonable as the VaR at the same duration and level is the barrier for returns to be included in the computation of the TCE. Finally, regression shows no significant link between the prediction error of the TCE and either the duration or the level or both. \\

\section{Conclusions and further directions}

\noindent For the eight different markets/indices under consideration, the VaR gives predictions that are more conservative than actual returns at the 10 day- 90\% and 20 day- 90\% levels. It is most accurate, when considering absolute error, at the 250 day-90\% level. The TCE is most conservative at the 250 day-95\% level, while it is most accurate at the 250 day-90\% level. Thus, we believe that the 10 day-90\% VaR and the 20 day-90\% VaR should be used in conjunction with the 250 day-90\% TCE and the 250 day-95\% TCE in determining a robust risk profile if our eight indices are the ones to be considered.\\

\noindent We anticipate further work on other commodities and on other stock indices from around the world to see if the trends spotted here are true in general. We would like to look particularly at emerging markets to see if there is some difference between these markets and more established ones. We would like to look at other commodity and equity markets to see if there is a great disconnect among risk measures in different markets or if the results of this paper are consistent across many markets. Finally, we plan to look at returns of individual stocks, both large and small caps, to see if the effect of aggregating stocks into an index has an effect on risk measures has the effect that would be expected by the Central Limit Theorem.\\

\end{document}